\documentclass[12pt,aps,prb,preprint]{revtex4} 
\usepackage{amsmath} 
\usepackage{amsfonts} 
\usepackage{amssymb}
  
\begin{document}

\title{The Spin-Statistics Connection In Classical Field Theory}

\author{J. A. Morgan}
\email{john.a.morgan@aero.org}
\affiliation{The Aerospace Corporation, P.\ O.\ Box 92957, Los Angeles,
California 90009}

\begin{abstract}
The spin-statistics connection is obtained for a simple formulation of a classical field
theory containing even and odd Grassmann variables.  To that end, the construction of irreducible 
canonical realizations of the rotation group corresponding to general causal fields is
reviewed.  The connection is obtained by imposing local commutativity on the fields and
exploiting the parity operation to exchange spatial coordinates in the scalar product of 
classical field evaluated at one spatial 
location with the same field evaluated at a distinct location.
The spin-statistics connection for irreducible canonical
realizations of the Poincar\'{e} group of spin $j$ is obtained in the form:
Classical fields and their conjugate momenta satisfy fundamental field-theoretic Poisson bracket 
relations for 2$j$
even, and fundamental Poisson antibracket relations for 2$j$ odd.

\end{abstract}

\maketitle

\section{Introduction}

Few would dispute that the field concept has been one of the happiest conceptual innovations in the 
history of 
physics.  From its nineteenth-century origins in the work of Faraday and Maxwell, to modern theories 
of fluid physics, the classical theory of elastic solids, 
relativistic gravitation, or of quantum field theory, the notion of a physical quantity which takes 
on values throughout a plenum has proven to be protean and endlessly fruitful.  Consider the success 
of the very prototype of a field theory, classical electrodynamics, in describing the propagation of 
electromagnetic radiation, in identifying that radiation with light, 
and in accounting for the transport of electromagnetic 
energy by light with the Poynting vector.  Imagine the difficulty of describing any of these 
phenomena without the aid
of electromagnetic fields.  Nor is the future development of quantum-mechanical descriptions of 
nature likely to abandon the field concept:
Whatever form a final theory of physics takes, whether cast in terms 
strings or branes, or other entities, "effective" quantum field theories valid at low energies-the 
definition of "low" varying, as occasion demands-will remain indispensible aids to 
practical calculation.

This paper presents a proof of the connection between spin and statistics for classical 
Grassmann fields.  An earlier paper~\cite{Morgan2004} presented 
a classical analog of the spin-statistics connection for pseudomechanics, a version of analytical 
dynamics containing even and odd Grassmann 
variables.~\cite{Casalbuoni1976,Casalbuoni1976a,BM1977,GalvaoTeitelboim1980,
Gomis1986a,DiVecchia1979}  
The method of proof relies on the canonical formalism for fields, suitably extended to include 
odd Grassmann variables.  Classical field theory is readily cast in terms of the canonical 
formalism,~\cite{Goldstein1950,SaletanCromer1971,SMukunda1974c} most familiarly as part of many
an introductory account of quantum field theory.~\cite{Wentzel1949,ItzyksonZuber1980} (But see
Refs.~\onlinecite{Weinberg1964,Weinberg1995nowt} for a different account of the foundations of 
quantum field theory.)  The literature on canonical formulations of classical field theory,
studied in their own right, is too large and various for a capsule summary.  A sampler of this 
body of research may be gleaned from
reviews in \cite{MarsdenEtAl1986,Kastrup1983,Salmon1988,JackiwEtAl2004}.  
On the other hand, apart from investigations inspired by 
supersymmetry~\cite{Kupershmidt1985,Kupershmidt1986}, 
classical treatments 
of odd  
Grassmann fields do not appear to be common.  Examples may be found in 
papers by Gozzi \textit{et al}~\cite{GozziEtAl1990} and Floreanini and 
Jackiw.~\cite{FloreaniniJackiw1988}

We begin by extending the canonical formalism for fields to classical Grassmann variables.
The classical equivalent of fermionic exchange symmetry appears in the properties of anticommuting 
Grassmann variables.  
The construction of irreducible canonical realizations for massive
fields possessing definite intrinsic spin, starting 
from the Lie algebra of the Poincar\'{e} group, is outlined next.  Irreducible
canonical realizations of the Poincar\'{e} group are classified in the same manner
as irreducible unitary representations in quantum field theory.  Canonical equivalents  
exist for the elements of the theory of unitary
representations in a Hilbert space, including ladder and Casimir operators.  
Of particular importance for the present problem, the canonical space
inversion (or parity) operation 
$\hat{P}$, and its action, are defined.
Finally, the spin-statistics connection is obtained by using the parity operation to exchange 
spatial coordinates in the scalar product of a field
evaluated at one spacetime location with the same field, evaluated at a distinct spacetime
location lying at spacelike interval from the first.

Local Poincar\'{e}
symmetry contributes three elements to the proof: (1) Local commutativity, (2) the properties of the 
rotational subgroup,
specifically the properties of irreducible canonical realizations of spin
degrees of freedom, and (3) the action of the discrete symmetry 
of parity transformation $\hat{P}$.  As in other proofs of the spin-statistics connection, invocation of 
Poincar\'{e} symmetry furnishes a sufficient condition for the connection. Its necessity is
addressed in Section \ref{section:Comments}.

Lower case Greek letters denote
either even or odd Grassmann fields, unless otherwise indicated in the text.
When it is desirable to distinguish 
even fields, they will be labelled by Latin letters of either case. 
Lower case Latin letters are also used for coordinates of spacetime
locations; $x \equiv ({\bf x},t)$ distinguishes space and time coordinates once a
spacelike foliation has been established.
Except when serving for spin degrees of freedom, lower case Greek indices run from 0 
to 3, while
lower case Latin indices run from 1 to 3.  
The summation convention applies
to repeated indices, unless the summation sign is explicity shown for emphasis.  When 
required for notational compactness, the partial derivative 
of $\psi_{\alpha}$ with respect to $x^{\mu}$ is written
\begin{equation}
\psi_{\alpha,\mu} \equiv {\partial{\psi_{\alpha}} \over \partial{x^{\mu}}}.
\end{equation}

\section{Classical Grassmann fields}

The commutation properties of Grassmann variables permit the realization of ferminonic and
bosonic exchange symmetry in a classical 
setting.~\cite{Morgan2004,Berezin1966,Swanson1992}  The familiar c-number bosonic 
variables
of traditional classical physics are \emph{even} Grassmann variables.  \emph{Odd} classical Grassmann
variables anticommute in a form of the exclusion principle.
A set of $n$ odd real Grassmann fields obeys the
anticommutation relations
\begin{equation}
\xi_{\mu}(x)\xi_{\nu}(y)+\xi_{\nu}(y)\xi_{\mu}(x)=0. \qquad (\mu,\nu \leq n)
\end{equation}
It follows that 
\begin{equation}
\xi^{2}_{\mu}(x)=0.
\end{equation}
One even and one odd variable,
in either order, commute. 
Differentiation on Grassmann variables can act from the right or the left. The
sign of the derivative of a product, for example, can depend on which derivative
is taken. Left differentiation, in accord with the convention in
Ref.~\onlinecite{Berezin1966}, is used exclusively in the following.
 
The formalism for
analytical dynamics of Grassmann variables, developed by Casalbuoni, Pauri, Prosperi, and Loinger, 
that found use in
Ref.~\onlinecite{Morgan2004}, can largely be translated directly into field-theoretic 
language.~\cite{SaletanCromer1971,SMukunda1974c,Salmon1988,Kupershmidt1992}  
Canonical transformations, for example, are defined in a manner that 
corresponds closely to analytical dynamics, as automorphisms of the fields that preserve Poisson
brackets. The necessary alterations to the definition of Poisson brackets for fields are given below.
The discussion will be limited to unconstrained systems for simplicity.

Fields are taken to be (Poincar\'{e}-symmetric) elements of the Hilbert space of complex square -
integrable functions.  The canonical realizations of the Poincar\'{e} group to be described below
form invariant subspaces of the Hilbert space.    
Fields are assumed to be "massive," so that it makes sense to speak of a rest frame for
them, \emph{i. e.,} a frame in which the spatial components of the four-momentum density 
vanish.~\cite{footnote1} 
The support of the
fields is taken to be a large region of Minkowski spacetime $R$ with boundary $\partial R$ on which 
normal gradients of the fields vanish. 
Let $q_{i}(x^{\mu})$, 
$i=1,\cdots , m$, be even field variables, and $\xi_{\alpha}(x^{\mu})$, 
$\alpha=1,\cdots , n$, be odd field variables. Given a Lagrangian 
\begin{equation}
L=L(q_{i},\xi_{\alpha},{\partial q_{i} \over \partial x^{\mu}},
{\partial \xi_{\alpha} \over \partial x^{\mu}}),
\end{equation}
define the Lagrangian density by
\begin{equation}
L=\int d^{3}x \mathcal{L}(q_{i},\xi_{\alpha},
{\partial q_{i} \over \partial x^{\mu}},{\partial \xi_{\alpha} \over \partial x^{\mu}}).
\end{equation}
The Euler-Lagrange equation for a field $\psi$ is
\begin{equation}
{\delta L \over \delta \psi} = 0
\end{equation}
with
\begin{equation}
{\delta L \over \delta \psi} \equiv 
{\partial \mathcal{L} \over \partial \psi}
-{\partial \over \partial x^{\mu}} {\partial \mathcal{L} \over \partial \psi_{,\mu}}
\end{equation}
for $\psi$ even or odd.
Generalized momenta are defined by
\begin{equation}
p^{i}={\delta L \over \delta q_{i,0}} \;\;\;\; 
\pi^{\alpha}={\delta L \over \delta \xi_{\alpha ,0}}.
\end{equation}
The Hamiltonian is given by
\begin{equation}
H=\int d^{3}x \mathcal{H}
\end{equation}
with
\begin{equation}
\mathcal{H} \equiv q_{i,0}p^{i}+\xi_{\alpha,0}\pi^{\alpha}-\mathcal{L}
\end{equation}
and Hamilton's equations are
\begin{eqnarray}
q_{i,0}={\delta H \over \delta p^{i}}\qquad
p^{i}_{,0}&=&-{\delta H \over \delta q_{i}}\\
\xi_{\alpha,0}=-{\delta H \over \delta \pi^{\alpha}} \qquad 
\pi^{\alpha}_{,0}&=&-{\delta H \over \delta \xi_{\alpha}}.
\end{eqnarray}

By way of introducing the Poisson bracket for 
Grassmann fields, let  
\begin{equation}
F=\int d^{3}x \mathcal{F}(q_{i},q_{i,k},\xi_{\alpha},\xi_{\alpha,k},p^{i},\pi^{\alpha})
\end{equation}
be an even functional and consider its total rate of change:
\begin{equation}
{d F \over d t} =  \int d^{3}x {\partial \mathcal{F} \over \partial t} 
+\int d^{3}x \Big \{  
{\partial \mathcal{F} \over \partial q_{i}} q_{i,0} +
{\partial \mathcal{F} \over \partial q_{i,k}} q_{i,k 0} +
{\partial \mathcal{F} \over \partial p^{i}} p^{i}_{,0} +
\xi_{\sigma,0} {\partial \mathcal{F} \over \partial \xi_{\sigma}} +
\xi_{\sigma,k 0} {\partial \mathcal{F} \over \partial \xi_{\sigma,k}} +
\pi^{\sigma}_{,0} {\partial \mathcal{F} \over \partial \pi^{\sigma}} \Big \}. \label{eq:dFdt}
\end{equation}
Consider the contribution from derivatives with respect to the odd field $\xi_{\mu}$
\begin{equation}
\int d^{3}x \Big \{ \xi_{\sigma ,0}{\partial \mathcal{F} \over \partial \xi_{\sigma}} +
 \xi_{\sigma,k 0} {\partial \mathcal{F} \over \partial \xi_{\sigma,k}}  \Big \} \label{eq:dFdtterm}
\end{equation}
We may write
\begin{equation}
{\partial \over \partial x_{k}} \Big 
[ \xi_{\sigma,0} {\partial \mathcal{F} \over \partial \xi_{\sigma,k}}  \Big ]
=\xi_{\sigma, 0} {\partial \over \partial x_{k}} \Big 
 [{\partial \mathcal{F} \over \partial \xi_{\sigma,k}} \Big ] 
+ \xi_{\sigma,k 0} {\partial \mathcal{F} \over \partial \xi_{\sigma,k}} 
\end{equation}
by virtue of the integrability condition
\begin{equation}
\psi_{\sigma,k 0}=\psi_{\sigma,0 k}
\end{equation}
thus obtaining for the rightmost term in (\ref{eq:dFdtterm})
\begin{equation}
\int d^{3}x \xi_{\sigma,k 0} {\partial \mathcal{F} \over \partial \xi_{\sigma,k}}  = 
\int d^{3}x \Big \{ {\partial \over \partial x_{k}} \Big 
[ \xi_{\sigma,0} {\partial \mathcal{F} \over \partial \xi_{\sigma,k}}  \Big ]
- \xi_{\sigma,0} {\partial \over \partial x_{k}} \Big 
[ {\partial \mathcal{F} \over \partial \xi_{\sigma,k}} \Big ] \Big \}.
\end{equation}
Converting the integral over all space of a total divergence to a surface integral with 
Gauss' theorem, and discarding the
surface term on $\partial R$ in the usual manner (a form of integration by parts) gives
\begin{equation}
\int d^{3}x \xi_{\sigma,0} \Big \{ {\partial \mathcal{F} \over \partial \xi_{\sigma}} 
-{\partial \over \partial x_{k}} \Big 
[ {\partial \mathcal{F} \over \partial \xi_{\sigma,k}} \Big ] \Big \}   
= \int d^{3}x \xi_{\sigma,0} {\delta F \over \delta \xi_{\sigma}}.
\end{equation}
Upon performing the same operation for the even fields, (\ref{eq:dFdt}) becomes
\begin{equation}
{d F \over d t}=\int d^{3}x  {\partial \mathcal{F} \over \partial t}
+\int d^{3}x \Big \{ 
{\delta F \over \delta q_{k}}  q_{k,0} 
+{\delta F \over \delta p^{k}} p^{k}_{,0}
+ \xi_{\sigma ,0} {\delta F \over \delta \xi_{\sigma}} 
+ \pi^{\sigma}_{,0} {\delta F \over \delta \pi^{\sigma}} \Big \}.
\end{equation}
Inserting Hamilton's equations for the derivatives of the field variables,
\begin{equation}
{d F \over d t}=\int d^{3}x {\partial \mathcal{F} \over \partial t}
+\int d^{3}x \Big \{ 
{\delta F \over \delta q_{i}} {\delta H \over \delta p^{i}}
-{\delta F \over \delta p^{i}} {\delta H \over \delta q_{i}}
-{\delta H \over \delta \pi^{\alpha}} {\delta F \over \delta \xi_{\alpha}} 
-{\delta H \over \delta \xi_{\alpha}} {\delta F \over \delta \pi^{\alpha}} 
 \Big \}. \label{eq:dFdTbrack}
\end{equation}
Rearranging (\ref{eq:dFdTbrack}) gives the template for the Poisson bracket for two even
functionals of the field variables~\cite{Casalbuoni1976,Casalbuoni1976a}
\begin{eqnarray}
[F,G]=\int d^{3}x \Big \{ {\delta F \over \delta q_{i}} {\delta G \over \delta
p^{i}} - {\delta G \over \delta q_{i}} {\delta F \over \delta p^{i}}
\Big \} +
\int d^{3}x \Big \{ {\delta F \over \delta \xi_{\alpha}} {\delta G \over \delta
\pi^{\alpha}} - {\delta G \over \delta \xi_{\alpha}} {\delta F \over
\delta \pi^{\alpha}} \Big \} \\
 =-[G,F].  \nonumber \label{eq:evenbracket} 
\end{eqnarray}
As in Ref.~\onlinecite{Casalbuoni1976}, the definition of the remaining brackets is fixed by 
requiring that multiplication of fields by an 
odd Grassmann constant give an algebra over the ring of Grassmann fields.
The bracket of two odd functionals $\theta$ and $\psi$ is given by
\begin{eqnarray}
[\theta,\psi]=\int d^{3}x \Big \{ {\delta \theta \over \delta q_{i}} {\delta \psi
\over 
\delta p^{i}} +
{\delta \psi \over \delta q_{i}} {\delta \theta \over \delta p^{i}}
\Big \} -
\int d^{3}x \Big \{ {\delta \theta \over \delta \xi_{\alpha}} {\delta \psi \over
\delta 
\pi^{\alpha}} +
{\delta\psi \over \delta \xi_{\alpha}} {\delta \theta \over \delta
\pi^{\alpha}} 
\Big \} \\
=[\psi,\theta], \nonumber
\end{eqnarray}
and is called an antibracket. When it is desired to emphasize the difference
between brackets of two even variables and antibrackets, they will be written
as $[F,G]^{-}$ and
$[\theta,\psi]^{+}$, respectively. For an odd and an even functional,
\begin{equation}
[\theta,F]=\int d^{3}x \Big \{ {\delta \theta \over \delta q_{i}} {\delta F \over
\delta p^{i}} - {\delta F \over \delta q_{i}} {\delta \theta \over
\delta p^{i}} \Big \} -
\int d^{3}x \Big \{ {\delta \theta \over \delta \xi_{\alpha}} {\delta F \over
\delta \pi^{\alpha}} + {\delta F \over \delta \xi_{\alpha}} {\delta
\theta \over \delta \pi^{\alpha}} \Big \} 
\end{equation}
The bracket between an even and an odd functional is defined so that
\begin{equation}
[F,\theta]=-[\theta,F].
\end{equation}

Casalbuoni\cite{Casalbuoni1976} 
has shown that the set of Poisson brackets and antibrackets in pseudomechanics
comprises a graded Lie algebra.\cite{BerezinKac1970,Corwin1975}
The development just given recapitulates that construction in terms of field-theoretic brackets.
Let $\varsigma_{f}=0$ for an even field and  $\varsigma_{\phi}=1$ for odd.  
The corresponding generalization of the Jacobi identity,\cite{Casalbuoni1976, Casalbuoni1976a,
BerezinKac1970, Corwin1975, Weinberg2000} 
\begin{equation}
(-1)^{\varsigma_{\pi}\varsigma_{\gamma}}[\gamma,[\rho,\pi]]+
(-1)^{\varsigma_{\gamma}\varsigma_{\rho}}[\rho,[\pi,\gamma]]+
(-1)^{\varsigma_{\rho}\varsigma_{\pi}}[\pi,[\gamma,\rho]]=0, \label{eq:Jaobi}
\end{equation}
required for the construction of canonical angular momentum ladder operators in
Sec.~\ref{ang.p}, is cumbersome to prove using elementary methods.  Given the validity of
(\ref{eq:Jaobi}) in pseudomechanics, its validity for classical fields follows from results 
obtained by Kupershmidt in a treatment of classical fields 
possessing Hamiltonian structure, using a geometric formulation of the calculus of 
variations.~\cite{Kupershmidt1985,Kupershmidt1986,Kupershmidt1992}
Jacobi identities for classical fields are not guaranteed to vanish under all 
circumstances.  However, the identities do vanish modulo a trivial divergence, whose 
contribution in the present discussion 
vanishes 
upon 
integration over $R$, 
by virtue of 
the assumption that all gradients vanish on $\partial R$.   

Fundamental Poisson brackets may be computed by making the substitution~\cite{SCromerp291}
\begin{equation}
\xi({\bf x},t)=\int{d^{3}z} \, \xi({\bf z},t) \delta ({\bf z}-{\bf x}),
\end{equation}
allowing one to regard $\xi$ formally as a functional for the purpose of permitting differentiation 
under the integral sign.  For two even fields we find:
\begin{equation}
[ q_{\mu}(t,{\bf x}),q_{\nu}(t,{\bf y})]^{-}=[p^{\mu}(t,{\bf x}),p^{\nu}(t,{\bf y}) ]^{-}
=0. 
\end{equation} 
The bracket between a field and its corresponding canonical momentum is computed with aid of the
identity~\cite{DiracDeltaID}
\begin{equation}
\delta({\bf x}-{\bf y})=\int{d^{3}z} \, \delta({\bf x}-{\bf z})\delta({\bf z}-{\bf y}),
\end{equation}
leading to 
\begin{equation}
[q_{\mu}(t,{\bf x}),p^{\nu}(t,{\bf y})]^{-}=\delta({\bf x}-{\bf y}) \delta_{\mu}^{\nu}.
\end{equation}
For odd fields and momenta, 
\begin{equation}
[ \xi_{\mu}(t,{\bf x}),\xi_{\nu}(t,{\bf y})]^{+}=[\pi^{\mu}(t,{\bf x}),\pi^{\nu}(t,{\bf y}) ]^{+}
=0. 
\end{equation} 
and
\begin{equation}
[\xi_{\mu}(t,{\bf x}),\pi^{\nu}(t,{\bf y})]^{+}=-\delta({\bf x}-{\bf y}) \delta_{\mu}^{\nu}.
\end{equation}

\section{Irreducible Canonical Realizations of the Poincar\'{e} Group}

The classical model of a field used in this paper is the canonical field-theoretic counterpart
of an irreducible representation of the Poincar\'{e} group used to describe a particle in 
quantum field theory.\cite{Wigner1939,Weinberg1995} Instead of commutation relations amongst
matrix generators of infinitesimal Lorentz transformations and rotations, one manipulates
Poisson brackets relating infinitesimal generators of canonical transformations.
Likewise, a functional on phase space which depends
solely upon the generators of the Lie algebra and which is an invariant
in all realizations of the Lie group is called a Casimir invariant, or simply a Casimir. 
Casimirs serve as the canonical equivalents of quantum-mechanical Casimir
operators.

Pauri and Prosperi\cite{PauriProsperi1966} developed the theory of canonical realizations of 
Lie groups, and later presented the canonical realization of the Poincar\'{e} group in 
detail.~\cite{PauriProsperi1975} (\emph{Vide.} also Ref.~\onlinecite{Dirac1949}.)  
Ref.~\onlinecite{Morgan2004} 
gives the extension of that analysis to canonical realizations in pseudomechanics.  
Classical fields  
of the simple kind considered in this paper satisfy Poisson bracket relations identical to those for
generalized coordinates and momenta in pseudomechanics.  Insofar as the development in 
Ref.~\onlinecite{PauriProsperi1975} depends only on algebraic relations of generators of 
canonical transformations and Poisson brackets, the results of that analysis can be translated
into the corresponding field-theoretic results.   
 
\subsection{The Poincar\'{e} group}
The realization of the Poincar\'{e} group in particle mechanics as a set of canonical 
transformations is presented in Ref.~\onlinecite{Morgan2004}.  
The corresponding development for fields is broadly analogous.
The effect of a general 
inhomogeneous Lorentz transformation $(\Lambda,a)$ on a 
four-vector $x_{\mu}$ is
\begin{equation}
x^{\prime\mu}=\Lambda^{\mu}_{\alpha}x^{\alpha}+a^{\mu}. \label{eq:inhomLorentz}
\end{equation}
If such a transformation is given a unitary representation (of any tensorial rank), the 
matrices $D$ of the representation
satisfy
\begin{equation}
D(\Lambda_{2},a_{2})D(\Lambda_{1},a_{1})
=D(\Lambda_{2}\Lambda_{1},a_{2}+\Lambda_{2}a_{1}). \label{eq:homomorphism}
\end{equation}
Sufficiently near to the origin,
\begin{eqnarray}
\Lambda^{\mu}_{\alpha}&=&\delta^{\mu}_{\alpha}+\omega^{\mu}_{\alpha}+O(\omega^{2})
\label{eq:Lx} \\
a^{\mu}&=&\epsilon^{\mu}, \label{eq:Px}
\end{eqnarray}
with $|\omega|$ and $|\epsilon| \ll 1$.
The quantity $\omega$ is antisymmetric in
its indices. To first order, a representation of (\ref{eq:Lx}) and (\ref{eq:Px}) is
\begin{equation}
D(1+\omega,\epsilon)=1+\frac{1}{2}\omega_{\mu\nu}\mathcal{M}^{\mu\nu}-
\epsilon_{\rho}\mathcal{P}^{\rho}
\end{equation}
where for each pair $(\mu,\nu)$ $\mathbf{\mathcal{M}}^{\mu\nu}=-\mathbf{\mathcal{M}}^{\nu\mu}$ is a 
matrix generator of generalized rotations, and 
$\mathbf{\mathcal{P}}^{\mu}$ is a matrix generator of translations.

The commutation relations relating
$\mathbf{\mathcal{M}}$ and $\mathbf{\mathcal{P}}$ comprise the Lie algebra of the Poincar\'{e} 
group.\cite{SMukunda1974,Weinberg1995,Dirac1949,Weinberg1972}
In classical field theory,  one may exploit the matrix representation
directly as in quantum field theory, or one may regard the commutation relations of the generators as determining Lie bracket
relations for an abstract representation of the Poincar\'{e} group.  We follow
the latter path here, and obtain a canonical
realization by replacing Lie brackets with Poisson brackets for the generators $M$ and $P$ of 
infinitesimal canonical
transformations corresponding to $\mathbf{\mathcal{M}}$ and $\mathbf{\mathcal{P}}$, 
as described in Ref.~\onlinecite{Morgan2004}.   

The ten generators of infinitesimal transformations for the Poincar\'{e} group
can be sorted into one temporal translation, three spatial translations, three
boosts, and three 
rotations.\cite{Weinberg1995,Dirac1949,Weinberg1972,PauriProsperi1975} 
Of these, only the generators of rotations and boosts, 
\begin{eqnarray}
\mathcal{J}_{i}&=&\epsilon_{ijk}\mathcal{M}^{jk} \qquad \mathcal{K}_{i}=\mathcal{M}^{i0} \qquad 
\mbox{$(i,j,k=1$--3}) ,
\end{eqnarray}
with commutators
\begin{eqnarray}
\mathcal{J}_{i} \mathcal{J}_{j}-\mathcal{J}_{j} \mathcal{J}_{i}& =  \label{eq:comJnJ}
& \epsilon_{ijk} \mathcal{J}_{k} \\
\mathcal{K}_{i} \mathcal{K}_{j}-\mathcal{K}_{j} \mathcal{K}_{i}& =
& -\epsilon_{ijk} \mathcal{J}_{k} \\
\mathcal{J}_{i} \mathcal{K}_{j}-\mathcal{K}_{j} \mathcal{J}_{i}& =  \label{eq:comJnK}
& \epsilon_{ijk} \mathcal{K}_{k} 
\end{eqnarray}
play a role in the subsequent development.

A field-theoretical realization of the Poincar\'{e} algebra can be constructed in more 
than one way;~\cite{footnote2} for 
illustrative purposes consider generators of 
canonical transformations derived from irreducible representations of the Poincar\'{e} group
that are bilinear in the fields and canonical momenta.~\cite{ItzyksonZuberB,SMukundaCh20}  
Given a matrix generator $\mathcal{G}_{\mu}(\mathbf{x},\mathbf{\nabla}_{x})$ of infinitesimal
transformations for an irreducible representation of the Poincar\'{e} group, 
one may obtain a canonical generator $G_{\mu}$ in the form
\begin{equation}
G_{\mu}=\int d^{3}x \, ( p^{i}\mathcal{G}_{\mu} q_{i} \label{eq:canonicalgen}
-\pi^{\alpha} \mathcal{G}_{\mu} \xi_{\alpha} )
\end{equation}
As in pseudomechanics, the canonical realization of
the Poincar\'{e} group possesses no nontrivial neutral elements.  The Poisson
bracket relations of the canonical realization are thus identical to those of the Lie algebra.  
The canonical generators
obtained by inserting (\ref{eq:comJnK}) into (\ref{eq:canonicalgen}) satisfy the relations
\begin{eqnarray}
{[}J_{i},J_{j}{]}^{-}& =& \epsilon_{ijk} J_{k} \label{eq:JnJ} \\ 
{[}K_{i},K_{j}{]}^{-}& =& -\epsilon_{ijk} J_{k} \\
{[}J_{i},K_{j}{]}^{-}& =& \epsilon_{ijk} K_{k}. \label{eq:JnK}
\end{eqnarray}
The realization of the Poincar\'{e} group so obtained consists of real generators.

The effect of an infinitesimal canonical transformation induced by a Lorentz transformation is given 
by
\begin{eqnarray}
\delta \xi_{\mu}=\delta \alpha_{\rho}[G_{\rho},\xi_{\mu}] \label{eq:infCTxi} \\
\delta \pi^{\mu}=\delta \alpha_{\rho}[G_{\rho},\pi^{\mu}] \label{eq:infCTpi}
\end{eqnarray}
where $G_{\rho}$ is one of the Lorentz generators and the  constants $\delta \alpha_{\rho}$ are the
Lorentz transformation parameters.
The effect of a finite transformation on a field is given formally by exponentiating the action of 
(\ref{eq:infCTxi})~\cite{CurrieEtAl1963}:
\begin{equation}
\xi'_{\mu}(x'_{\nu})=
exp \left (\alpha_{\rho}[ G_{\rho},...] \right ) \xi_{\mu}(x_{\lambda}) 
\end{equation}
In the case of an irreducible realization, the result may 
also be expressed as
\begin{equation}
\xi'_{\mu}(x'_{\nu})=D_{\mu}^{\rho}(\Lambda)\xi_{\rho}(\Lambda^{\lambda}_{\nu}
 x_{\lambda}+a_{\nu})
\end{equation}
where the matrix $D_{\mu}^{\rho}(\Lambda)$ belongs to the corresponding irreducible representation 
of the Lorentz group.~\cite{SMukundaDmat}

\subsection{Angular momentum and irreducible canonical realizations}\label{ang.p}

The behavior under rotations of a field that transforms as an irreducible canonical realization
with a definite angular momentum closely resembles that of a corresponding quantum-mechanical
system.
Canonical angular momentum variables satisfy bracket relations that
form a subalgebra of the Poincar\'{e} algebra decoupled from the boost degrees of freedom.
As a result, the spin degrees of freedom of a massive field in its rest frame are 
treated in the same way as in nonrelativistic quantum mechanics.  

Suppose that a field is an irreducible canonical realization possessing definite angular 
momentum.  Fix the direction of the $z$-axis along 
its spatial part and write $\xi$ for the magnitude.  The change in the value of $\xi$ induced by an 
infinitesimal rotation about $z$ is
\begin{equation}
\xi(x)\Rightarrow\xi(x)+\delta\phi[\xi,J_{z}].
\end{equation}
If $\xi$ has rotational symmetry about the axis defining
$\phi$, the effect of this transformation must be equivalent to
multiplication by a phase:
\begin{equation}
\xi\Rightarrow\xi+im\delta\phi\xi,
\end{equation}
or
\begin{equation}
[J_{z},\xi]=-im\xi \label{eq:Jzbracket}
\end{equation}
One may regard this relation as a kind of 
eigenvector condition, and label $\xi$ by its eigenvalue $m$ as $\xi_{m}$.~\cite{Loinger1963}
The angular momentum ladder operators 
\begin{equation}
J_{\pm}=J_{x} \pm iJ_{y},
\end{equation}
satisfy
\begin{eqnarray}
[J_{+},J_{-}]^{-}&=&-2iJ_{z}  \\
{[}J_{z},J_{\pm}]^{-}&=&\mp J_{\pm}. \label{eq:JzJpmbracket}
\end{eqnarray}

The quantity
\begin{equation}
J^{2}=J_{x}^{2}+J_{y}^{2}+J_{z}^{2} 
=J_{z}^{2}+\frac{1}{2}[J_{+}J_{-}+J_{-}J_{+}],
\end{equation}
has vanishing brackets with all the generators of rotations in an irreducible
realization. It is thus a Casimir which, in any irreducible
canonical realization, is a constant,\cite{SMukunda1974b} so
that
\begin{equation}
[J^{2},\xi_{m}]=\mbox{constant}\,\xi_{m} \equiv j(j+1)\xi_{m}.
\end{equation}
Irreducible realizations $\xi_{m}$ for definite $m$ are thus more properly labeled by both
eigenvalues $j$ and $m$ as $\xi_{jm}$.

The $z$-projection of the angular momentum of $[J_{\pm},\xi_{jm}]$ is obtained by computing
$[J_{z},[J_{\pm},\xi_{jm}]]$
with the aid of the Jacobi identity (\ref{eq:Jaobi})
and the bracket relations of the ladder operators from (\ref{eq:JzJpmbracket}):
\begin{equation}
[J_{z},[J_{\pm},\xi_{jm}]]=-i(m\pm 1)[J_{\pm},\xi_{jm}].
\label{this}
\end{equation}

Continuing in this way, we may obtain the action of the ladder operators in the
canonical formalism just as in quantum mechanics. In
particular\cite{Loinger1963},
\begin{equation}
[J_{\pm},\xi_{jm}]=-i\sqrt{(j\mp m)(j\pm m+1)}\xi_{jm\pm1}.
\label{eq:ladderopdef}
\end{equation}
It follows~\cite{Tung1985a,Loinger1963,Dirac1958} that, as in quantum mechanics, realizations
of integral and half-integral $j$ occur. The eigenvectors of 
$J_{z}$ given by (\ref{eq:Jzbracket}) have integer or half-integer eigenvalues 
$-j\leq m \leq j$,
and span a $2j+1$ dimensional invariant subspace of the Hilbert space of
canonical realizations of the rotation subgroup.
In what follows, the $\xi_{jm}$ will be treated collectively as components of an irreducible
spherical tensor of rank $j$.

\subsection{Classification of Irreducible Realizations of the Poincar\'{e} Group} \label{sect3c}

Irreducible canonical realizations of the Poincar\'{e} group may be classified in a manner
entirely analogous to the method used in quantum field theory.  Construct infinitesmal generators
from the quantities defined in (\ref{eq:JnJ}-\ref{eq:JnK})
\begin{equation}
\mbox{\bf{A}}=\frac{1}{2} (\mbox{\bf{J}}+i\mbox{\bf{K}}) \label{eq:defA}
\end{equation}
and 
\begin{equation}
\mbox{\bf{B}}=\frac{1}{2} (\mbox{\bf{J}}-i\mbox{\bf{K}}) \label{eq:defB}
\end{equation}
These satisfy the bracket relations:
\begin{eqnarray}
[A_{i},A_{j}]^{-}&=&\epsilon_{ijk}A_{k}  \qquad  [B_{i},B_{j}]^{-}=\epsilon_{ijk}B_{k}  \qquad 
[A_{i},B_{j}]^{-}=0.
\end{eqnarray}
The first two of these are identical to the Poisson brackets for $\mbox{\bf{J}}.$
The ladder operator formalism developed in the preceding section therefore may be used to generate 
the components of an irreducible canonical realization in exactly the same way that a complete set
of $m$-values is generated for spin $j$.  The irreducible canonical realizations are
classified by a pair of indices $(a,b)$, both of which may be either integral or half-integral.
Generators 
$A_{3}$ and $B_{3}$ act on irreducible canonical realizations $\xi^{(a,b)}_{kl}$ according to
\begin{equation}
[A_{3},\xi^{(a,b)}_{kl}]=-ik\xi^{(a,b)}_{kl}
\end{equation}
and 
\begin{equation}
[B_{3},\xi^{(a,b)}_{kl}]=-il\xi^{(a,b)}_{kl} 
\end{equation}
where the eigenvalues
\begin{equation}
 k = -a, -a+1, \cdots,+a
\end{equation}
and
\begin{equation}
l = -b, -b+1, \cdots,+b.
\end{equation}
A ladder operator for $\bf{A}$ is given by
\begin{equation}
[A_{1} \pm iA_{2},\xi^{(a,b)}_{kl}]=
-i\sqrt{(a\mp k)(a\pm k+1)}\xi^{(a,b)}_{k\pm1,l} \label{defApm}
\end{equation}
and similarly for $\bf{B}$.

Canonical realizations with a definite spin $j$ are constructed from combinations of the
$(a,b)$ realizations.  Thus, for example, 
a scalar field belongs to the $(0,0)$ irreducible
realization, while the Dirac field belongs to the $(\frac{1}{2},0) \oplus (0,\frac{1}{2})$ 
realization.  Each such realization spans a $(2a+1)(2b+1)$ subspace of the Hilbert space of
irreducible canonical realizations.  Realizations with integral $j$ are sometimes called
\emph{tensorial}, and those with half-integral $j$, \emph{spinorial}.~\cite{Weinberg1972}
Writing
\begin{equation}
J_{i}=A_{i}+B_{i},
\end{equation}
we see that $J_{3}$ will have eigenvalue $m \equiv k+l$ and     
$J^{2}$ will have eigenvalue $j(j+1)$ where $j=a+b$.  
Following Weinberg~\cite{WeinbergCF}, an element of the $(a,b)$ irreducible canonical
realization is called a general causal field.
In the following, once $(a,b)$ is fixed,
the field $\xi$ will be written
\begin{equation}
\xi_{jm}\equiv\xi^{(a,b)}_{kl} \label{eq:irrep}
\end{equation}
with
\begin{eqnarray}
j=a+b  \qquad  && m=k+l
\end{eqnarray}
or simply
\begin{equation}
\xi_{m}\equiv\xi_{jm}
\end{equation}
unless it is necessary to specify the exact irreducible canonical realization under discussion.  

\section{Parity and classical fields} \label{P}

\subsection{Canonical realizations of the parity operation}

The preceding Section treated the effect of continuous coordinate transformations upon 
classical Grassmann fields required for the proof of the 
spin-statistics connection.  A complete realization of the  Poincar\'{e} 
group also includes the discrete transformations of parity, 
time reversal, and charge conjugation.
Classical analogs of charge conjugation and time reversal are discussed 
in Ref.~\onlinecite{Morgan2004}.
The action
of parity on a scalar function $\xi$ of spacetime location $\mathbf{x},t$ 
is 
\begin{equation}
\mathcal{P}(\xi(\mathbf{x},t))=\eta \xi(-\mathbf{x},t).
\end{equation}

$\mathcal{P}$ commutes with the generators of time translations and rotations, but anticommutes with 
the generators of spatial translations and boosts.  Pauri and Prosperi~\cite{PauriProsperi1975} show
that the operator $\hat{P}$ that realizes $\mathcal{P}$ in a canonical realization of the 
Poincar\'{e} group has the action
\begin{equation}
\hat{P}(\mathbf{J})=\mathbf{J}  \label{PonJ}
\end{equation}
and
\begin{equation}
\hat{P}(\bf{K})=-\bf{K}  \label{PonK}
\end{equation}
on the generator of infinitesimal canonical transformations for rotations and boosts, respectively.
Equations (\ref{PonJ}) and (\ref{PonK}) are to be understood as shorthand for relations of the 
form
\begin{equation}
[\hat{P}(\mathbf{Q}),\xi]=\pm[\mathbf{Q},\xi],\forall~\xi.
\end{equation}
In particular (\emph{vide.} also (\ref{eq:etaspinindep}) below), 
the action of $\hat{P}$ is diagonal on components of a field of definite spin in 
a spherical tensor basis, allowing us to write:
\begin{equation}
\hat{P}( \xi_{m}(x))=\eta \xi_{m}(x') \label{eq:PonXi}
\end{equation}
where
\begin{eqnarray}
\bf{x}'=-\bf{x}  \qquad  &&  t'=t.
\end{eqnarray}
Because $\hat{P}^{2}=1$, 
\begin{equation}
\eta^2=1;
\end{equation}
\begin{equation}
\eta=\pm1 \label{eq:IntrinsicP}
\end{equation}
for any field.

In quantum mechanics, the parity of a state is a multiplicative quantum number.    
The classical statement of this property to be used in Section \ref{SST} is that 
the parity of the scalar product of two 
fields is the product of the parities of the individual fields.
Start with the observation that a scalar function of 
position has even parity under space inversion.  A scalar is unchanged by any transformation of
reference frame.~\cite{Weinberg1972b}  If one expresses a scalar field $\xi(\mathbf{r})$ in terms of
a new coordinate frame $\mathbf{r}''$ as $\xi''(\mathbf{r}'')$,\cite{Edmonds1960nowt} 
\begin{equation}
\xi''(\mathbf{r}'')=\xi(\mathbf{r}). \label{scalarstuff0}
\end{equation}
Fix a common origin for both $\mathbf{r}$ and $\mathbf{r}''$.  The frame $\mathbf{r}''$ is assumed 
to differ from $\mathbf{r}$ by
a rotation defined by Euler angles $\alpha,\beta,\gamma$.  Denote the 
action of 
the rotation that carries $\mathbf{r}$ to $\mathbf{r}''$ by the
rotation operator $\mathcal{D}^{(1)}(\alpha~\beta~\gamma)$ and that which carries $\xi$ to $\xi''$ by 
$\mathcal{D}^{(0)}(\alpha~\beta~\gamma)$:
\begin{equation}
\mathbf{r}''=\mathcal{D}^{(1)}(\alpha~\beta~\gamma)\mathbf{r}
\end{equation}
and
\begin{equation}
\xi''=\mathcal{D}^{(0)}(\alpha~\beta~\gamma)\xi.
\end{equation}
The rotation operator $\mathcal{D}^{(0)}(\alpha~\beta~\gamma)$ for a scalar is simply unity.
By (\ref{scalarstuff0}) we thus have
\begin{equation}
\xi(\mathcal{D}^{(1)}(\alpha~\beta~\gamma)\mathbf{r})=\xi(\mathbf{r})
\end{equation}
for any rotation $\alpha,\beta,\gamma$.  Once an origin has been fixed, the value of $\xi$ does not 
depend upon
the orientation of the coordinate axes.  Accordingly, $\xi$ can depend only upon the magnitude of
$\mathbf{r}$:
\begin{equation}
\xi(\mathbf{r})=\xi(|\mathbf{r}|).
\end{equation}
In particular, 
 \begin{equation}
\xi(-\mathbf{r})=\xi(\mathbf{r}). \label{evenness}
\end{equation}
The parity of a scalar field is thus $\eta=+1$.

Next, the scalar product of two fields with the same rank is given by\cite{Edmonds1960}
\begin{equation}
\xi \cdot \zeta =
\sum_{m}(-1)^{m}\xi_{m} \label{eq:scalarprod0}
\zeta_{-m}.  
\end{equation}
Equation~(\ref{eq:scalarprod0}) is proportional to the expression for the coupling of $\xi$ and
$\zeta$ to spin zero\cite{Racah1942}, \emph{i. e.}, a scalar function of position.
Consider the effect of the parity operation on the scalar product in (\ref{eq:scalarprod0}),
\begin{equation}
\hat{P} (\xi(x) \cdot \zeta(x)) = 
\eta_{\xi \zeta} \xi(x') \cdot \zeta(x').
\end{equation}
With the aid of (\ref{eq:PonXi}) and (\ref{eq:IntrinsicP}), this may be 
written
\begin{equation}
\eta_{\xi \zeta} \hat{P} (\xi(x) \cdot \zeta(x)) = 
\eta_{\xi} \eta_{\zeta} \hat{P}(\xi(x)) \cdot \hat{P}(\zeta(x)). \label{eq:P2fields}
\end{equation}
The only choice consistent with a nonvanishing scalar product is readily seen to be 
\begin{equation}
\hat{P}(\xi(x)\cdot\zeta(x))=
\hat{P}(\xi(x))\cdot\hat{P}(\zeta(x))	\label{sprodfakt1}
\end{equation}
and
\begin{equation}
\eta_{\xi \zeta}=\eta_{\xi} \eta_{\zeta}. \label{sprodfakt2}
\end{equation}

\subsection{Parity and general canonical realizations of the Poincar\'{e} group}

Consider next the effect of space inversion on the generators $\bf{A}$ and $\bf{B}$ of 
(\ref{eq:defA}) and (\ref{eq:defB}).  Recall from (\ref{PonJ}) and (\ref{PonK}) that 
under the parity operation the generator 
of rotations is even, and that of boosts is odd, 
with the result 
\begin{eqnarray}
\hat{P}(\bf{A})=\bf{B}  \qquad  &&  \hat{P}(\bf{B})=\bf{A}.
\end{eqnarray}
Making use of the identity 
\begin{equation}
\hat{P}([\xi,\zeta])=[\hat{P}(\xi),\hat{P}(\zeta)]
\end{equation}
which follows from the definition of a canonical transformation and the
action of parity~\cite{PauriProsperi1975,Kupershmidt1986b}, 
apply the parity operation to (\ref{defApm}) to obtain~\cite{WeinbergParity}:
\begin{eqnarray}
\hat{P}([A_{\pm},\xi^{(a,b)}_{kl}])=
-i\sqrt{(a\mp k)(a\pm k+1)}\hat{P}(\xi^{(a,b)}_{k\pm1,l})
\label{eq:PonbracketA} \\
=[\hat{P}(A_{\pm}),\hat{P}(\xi^{(a,b)}_{kl})] \\
=[B_{\pm},\hat{P}(\xi^{(a,b)}_{kl})]. 
\end{eqnarray}
By the same reasoning: 
\begin{equation}
[A_{\pm},\hat{P}(\xi^{(a,b)}_{kl})]=
-i\sqrt{(b\mp k)(b\pm k+1)}\hat{P}(\xi^{(a,b)}_{k,l\pm1}) \label{eq:PonbracketB}
\end{equation}
From (\ref{eq:PonbracketA}-\ref{eq:PonbracketB}) we conclude 
\begin{equation}
\hat{P}(\xi^{(a,b)}_{kl}(\mathbf{x},t)) \propto \xi^{(b,a)}_{lk}(-\mathbf{x},t) 
\end{equation}
Set $j=a+b$ and write 
\begin{equation}
\hat{P}(\xi^{(a,b)}_{kl}(\mathbf{x},t)) \equiv 
\eta_{kl} \Phi(j,a,b) \xi^{(b,a)}_{lk}(-\mathbf{x},t) \label{eq:PcausalForm}
\end{equation}
with
\begin{equation}
\eta_{kl}^{2} \Phi^{2}(j,a,b)=1
\end{equation}
Then
\begin{eqnarray}
\hat{P}([A_{\pm},\xi^{(a,b)}_{kl}])=
-i\sqrt{(a\mp k)(a\pm k+1)}\eta_{k\pm1 l} \Phi(j,a,b) \xi^{(b,a)}_{lk\pm1}(-\mathbf{x},t) 
\label{eqn1} \\ 
=[B_{\pm},\eta_{kl} \Phi(j,a,b)\xi^{(b,a)}_{lk}(-\mathbf{x},t)] \\
=-i\sqrt{(a\mp k)(a\pm k+1)}\eta_{kl} \Phi(j,a,b)\xi^{(b,a)}_{lk\pm1}(-\mathbf{x},t). \label{eqn2}
\end{eqnarray}
Divide out common terms in (\ref{eqn1}) and (\ref{eqn2}) to obtain
\begin{equation}
\eta_{k\pm1 l}=\eta_{kl}
\end{equation}
and likewise upon exchanging $\bf{A}$ and $\bf{B}$ in the foregoing,
\begin{equation}
\eta_{k l\pm1}=\eta_{kl} \equiv \eta. \label{eq:etaspinindep}
\end{equation}
As before, $\hat{P}^{2} \equiv 1$ allows us to write $\eta^{2}=1$, so
\begin{equation}
\hat{P}(\xi^{(a,b)}_{kl}(\mathbf{x},t)) \equiv 
\eta \Phi(j,a,b) \xi^{(b,a)}_{lk}(-\mathbf{x},t). \label{eq:PcausalFinal}
\end{equation}
Although we make no use of it, the choice of $\Phi$ consistent with conventions for
general causal fields in quantum field theory~\cite{Weinberg1995a} is
\begin{equation}
\Phi(j,a,b) = (-1)^{(a+b-j)}.
\end{equation}

\section{Connection between spin and statistics} \label{SST}

Before attacking the case of general causal fields, the method of proof is worked out for the
simpler case of $(j,0)$ representations.~\cite{Weinberg1964}
Define the (Weinberg) field 
\begin{equation}
\xi_{m} \equiv \xi^{(j,0)}_{m0}
\end{equation}
where $m$ runs from $-j$ to $j$. It will be shown that imposing local commutativity on components of 
a field $\xi_{m}(\bf{x},t)$
leads to the spin-statistics connection.~\cite{Morgan2005} 

The central element of the proof relies on the scalar product of a certain field evaluated at one 
spacetime location with the same field, evaluated at a point lying at spacelike
interval from the first.  A Lorentz frame exists in which the two points occur at equal time.  The
fields may therefore be written as  
$\xi(\mbox{\bf{y+x}},t)$ and $\xi(\mbox{\bf{y}},t)$,  
and the scalar product as
\begin{equation}
\xi (\mbox{\bf{y+x}},t)  \cdot \xi (\mbox{\bf{y}},t) 
\end{equation}
By translational invariance, this must 
be identical with
\begin{equation}
\xi (\mbox{\bf{x}},t)  \cdot \xi (\mbox{\bf{0}},t) 
= \xi (\mbox{\bf{x}}/2,t)  \cdot \xi (\mbox{\bf{-x}}/2,t) 
\end{equation}
which we write as
\begin{equation}
\xi (\mbox{\bf{x}},t)  \cdot \xi (\mbox{\bf{-x}},t) 
\end{equation}
from here on.
While this object can be regarded as a purely formal
device, it is closely related to a quantity which finds use elsewhere is classical physics, the 
correlation function.  
The spatial autocorrelation of $\xi$
\begin{equation}
\mbox{\bf{g}}(\mbox{\bf{x}},t)=\langle \xi (\mbox{\bf{y+x}},t) \xi (\mbox{\bf{y}},t) \rangle
\end{equation}
is of great significance in theories of statistical fluctuations.~\cite{LandauLifshitzStat,Huang1987}
The angle brackets denote a spatial average over $\bf{y}$.  While correlation functions find more 
use in
classical statistical mechanics, where they are used to describe the relation between fluctuations in
particle occupation number at distinct points in (say) an ideal gas, correlation functions appear
in continuum physics as well, most notably in the theory of interference of electromagnetic 
fields,~\cite{BornWolf1980,MandelWolf1965} but also in descriptions of phenomena as diverse as 
intensity interferometry in radio and optical astronomy,~\cite{HBTwiss1954}
pressure fluctuations in
fluid mechanics~\cite{LandauLifshitzFM}, fluctuation-dissipation relations in acoustics and
electromagnetism~\cite{Weber1956,CallenWelton1951}, and critical opalescence~\cite{Domb}.  
Tensor correlation functions also find use in the study of anisotropy and polarization of the
cosmic microwave background.~\cite{NgLiu1997}

Commence by disposing of the possibility that the quantity
\begin{equation}
\xi (\mbox{\bf{x}},t) \cdot \xi (-\mbox{\bf{x}},t)  \label{eq:introsprod}
\end{equation}
might vanish identically for nontrivial fields.  We do so by constructing a field for which 
(\ref{eq:introsprod}) may be seen to be nonvanishing.  
From the scalar product of fields $\psi$ and $\phi$ of the same rank form 
\begin{equation}
(\psi,\phi) \equiv 
\frac{1}{2j+1}\int {d^{3}x} \, \psi^{*}(\mbox{\bf{x}},t) \cdot \phi(\mbox{\bf{x}},t). \label{eq:IP}
\end{equation}
Equation (\ref{eq:IP}) defines an inner product on the Hilbert space of complex
square-integrable functions on a spacelike slice of $R$:
(1) $(\psi+\eta,\phi)=(\psi,\phi)+(\eta,\phi)$
(2) $(a\psi,b\phi)$ is linear in $b$ and antilinear in $a$.  
(3) $(\psi,\phi)^{*}=(\phi,\psi)$
(4) $(\psi,\psi) \ge 0$
and $(\psi,\psi)$ vanishes iff $\psi$ does.  We may see this last as follows.  In
\begin{equation}
(\psi,\psi) = \frac{1}{2j+1}\int {d^{3}x} \, \psi^{*}(\mbox{\bf{x}},t) \cdot \psi(\mbox{\bf{x}},t) 
\label{eq:norm}
\end{equation}
we may expand the angular dependence of $\psi$ in normal modes.  The  spin-weighted spherical
harmonics$\,_{s}Y_{jm}$ generalize
ordinary spherical harmonics to arbitrary (to include half-integral) eigenvalues of 
dimensionless angular 
momentum.~\cite{NewmanPenrose1966,GoldbergEtAl1967,HuWhite1997}  The
spin-weight $s$ is the negative of a dimensionless helicity.~\cite{Campbell1971}
An irreducible realization $\psi$ of 
spin $j$ appearing in the integrand of equation (\ref{eq:norm}) may be written 
\begin{equation}
\psi_{m}(r,\Omega)=f_{j}(r)\,_{s}Y_{jm}(\Omega)
\end{equation}
at radius $r$ for some value of $s$.  Because the canonical 
ladder operators that raise and lower $m$ and $s$
act on angular degrees of freedom only, the radial weight can have no dependence upon either
$m$ or $s$.~\cite{Baym69}
The harmonics satisfy
\begin{equation}
\,_{s}Y^{*}_{jm}=(-1)^{(s+m)}\,_{s}Y_{jm}
\end{equation}
and
\begin{equation}
\int {d\Omega} \,_{s}Y^{*}_{jm'}\,_{s}Y_{jm}=\delta_{m'm}.
\end{equation}
It therefore may be arranged that
\begin{equation}
(\psi,\psi) = \int {r^{2}dr} \, f_{j}^{*}(r) f_{j}(r) \ge 0
\end{equation}
(One must watch the order of factors if $\psi$ is odd.)
In particular, if $\psi \cdot \psi^{*}=\psi^{*} \cdot \psi=0$ 
almost everywhere, then $(\psi,\psi)=0$. But $(\psi,\psi)=0$ iff $\psi=0$ a.e.  

Assume $\zeta$ is a nonvanishing field belonging to a spin $j$ irreducible canonical realization.
From $\zeta$ form
\begin{equation}
\xi_{m}(\mbox{\bf{x}},t)=\zeta^{*}_{m}(\mbox{\bf{x}},t) 
\pm \zeta_{m}(-\mbox{\bf{x}},t).  \label{eq:xidef}
\end{equation}
We have
\begin{equation}
\xi(\mbox{\bf{x}},t) \cdot \xi (-\mbox{\bf{x}},t)= 
\pm \xi(\mbox{\bf{x}},t) \cdot \xi^{*}(\mbox{\bf{x}},t). \label{eq:xisprods}
\end{equation}
As a general rule, the field $\xi$ will have nonvanishing norm and the RHS of (\ref{eq:xisprods})
will differ from zero.  But suppose that for one choice of sign in (\ref{eq:xidef}), $\xi$ 
were to vanish  $\forall \, \mbox{\bf{x}}$.
In that event $\xi$, and hence (\ref{eq:xisprods}), cannot vanish for the other choice.  
We suppose in what follows that the appropriate choice of sign has been made, if necessary, and 
that (\ref{eq:introsprod}) is therefore nonvanishing on some open set of $\mbox{\bf{x}}$.  

The effect of
$\hat{P}$ on the scalar product of $\xi({\bf x},t)$ and $\xi(-{\bf x},t)$ is, according to 
(\ref{eq:PcausalFinal}) for $a=b=0$, 
\begin{eqnarray}
 \hat{P}( \xi (\mbox{\bf{x}},t) \cdot \xi (\mbox{-\bf{x}},t)) \nonumber \\
=\hat{P} (\xi (\mbox{\bf{x}},t))  \cdot \hat{P} (\xi (\mbox{-\bf{x}},t)) \nonumber \\
=\xi (-\mbox{\bf{x}},t) \cdot \xi (\mbox{\bf{x}},t)\label{eq:sproduct1}
\end{eqnarray}
This quantity is the product of 
two terms with the same parity, and by (\ref{sprodfakt2}) is even parity itself.    
Considered as a function of $\mbox{\bf{x}}$, an even parity scalar obeys
$\hat{P} (f(\mbox{\bf{x}}))=f(\mbox{\bf{x}})$, thus we have
\begin{equation}
\xi (\mbox{\bf{x}},t) \cdot \xi (-\mbox{\bf{x}},t)
=\xi (-\mbox{\bf{x}},t) \cdot \xi (\mbox{\bf{x}},t).  \label{eq:sproduct2p5}
\end{equation}

Commutation relations of a causal field ($-$ for Bose, $+$ for Fermi) 
vanish outside the light cone:
\begin{equation}
\xi_{m}(\mbox{\bf{x}},t) \xi_{n}(-\mbox{\bf{x}},t)
 \pm \xi_{n}(-\mbox{\bf{x}},t) \xi_{m}(\mbox{\bf{x}},t)=0 \label{eq:ccr}
\end{equation}
By (\ref{eq:scalarprod0}),
\begin{equation}
\xi(-\mbox{\bf{x}},t) \cdot \xi (\mbox{\bf{x}},t)=
\pm\sum_{m}(-1)^{m}\xi_{-m}
(\mbox{\bf{x}},t)\xi_{m}(-\mbox{\bf{x}},t), \label{eq:sproduct4}
\end{equation}
for Bose ($+$) or Fermi ($-$) fields, respectively.  Invert the order of summation by replacing
$m$ with $-m'$: 
\begin{equation}
\xi(-\mbox{\bf{x}},t) \cdot \xi (\mbox{\bf{x}},t)=
\pm\sum_{m'}(-1)^{-m'}\xi_{m'}
(\mbox{\bf{x}},t)\xi_{-m'}(-\mbox{\bf{x}},t).  \label{eq:sproduct2p75}
\end{equation}
Now, $j+m'$ is always an integer, and $2j+2m'$ an even integer.  We may write
\begin{eqnarray}
(-1)^{-m'}=(-1)^{-m'}(-1)^{2j+2m'} \nonumber \\
=(-1)^{2j}(-1)^{m'}
\end{eqnarray}
in (\ref{eq:sproduct2p75}) to obtain for (\ref{eq:sproduct2p5})
\begin{equation}
\xi(\mbox{\bf{x}},t) \cdot \xi (-\mbox{\bf{x}},t)=
\pm(-1)^{2j}\xi (\mbox{\bf{x}},t) \cdot \xi (-\mbox{\bf{x}},t). \label{eq:sproduct5}
\end{equation}
or
\begin{equation}
1=\pm(-1)^{2j}. \label{eq:drumroll}
\end{equation}
Equation~(\ref{eq:drumroll}) is a statement of the connection between spin and statistics.

The extension of the argument just given to the case of the general $(a,b)$ representation 
is straightforward.  
The field $\xi^{(ab)}_{kl}$ now carries two indices $-a \le k \le a$ and
$-b \le l \le b$, and (\ref{eq:scalarprod0}) is replaced with an 
expresson that couples two $(a,b)$ spherical tensors to a $(0,0)$ scalar, in a generalization 
of Racah's~\cite{Racah1942} original derivation of (\ref{eq:scalarprod0}).  That expression now 
becomes (retaining the dot product notation)
\begin{eqnarray}
\sum_{kl} \left( \begin{array}{rlc}
a & a & 0 \\
-k & k & 0 
\end{array} \right)
\left( \begin{array}{rlc}
b & b & 0 \\
-l & l & 0 
\end{array} \right) 
\xi^{(a,b)}_{kl}(-\mbox{\bf{x}},t) \xi^{(a,b)}_{-k-l}(\mbox{\bf{x}},t) \nonumber \\
\propto \sum_{kl}(-1)^{m}\xi^{(a,b)}_{kl}(-\mbox{\bf{x}},t) 
\xi^{(a,b)}_{-k-l}(\mbox{\bf{x}},t) \nonumber \\
\equiv \xi(-\mbox{\bf{x}},t) \cdot \xi (\mbox{\bf{x}},t)  \label{eq:newsproduct}
\end{eqnarray}
where $m=k+l$, and the objects in parentheses are Wigner 3j symbols.  It is readily shown that
(\ref{sprodfakt1}) and (\ref{sprodfakt2}) are valid for the generalized scalar product, and
that (\ref{eq:newsproduct}) vanishes iff $\xi (\mbox{\bf{x}},t)$
does.
By (\ref{eq:PcausalFinal}) for the (0,0) realization, the result of applying $\hat{P}$ 
to  (\ref{eq:newsproduct}) once again gives (\ref{eq:sproduct2p5}).
Both the spin $j$ and summation index $m$ are half-integral iff one of 
$a$ and $b$ is half-integral.  Therefore, (\ref{eq:sproduct5}), and thus
(\ref{eq:drumroll}), hold for the general $(a,b)$ realization, as well.

We conclude classical 
fields which are irreducible canonical realizations of 
spin $j$ must be commuting, even Grassmann variables if $j$ is an integer, and anticommuting, 
odd Grassmann variables if $j$ is half-integral.  From the symmetry properties of brackets 
given earlier follows immediately the conclusion that irreducible canonical 
realizations for integral $j$ obey Poisson bracket relations, while realizations for half-integral 
$j$ obey Poisson antibracket relations.  If $\pi^{\mu}$ is 
the momentum conjugate to $\xi_{\mu}$, then the brackets are:
\begin{equation}
[ \xi_{\mu}(t,{\bf x}),\xi_{\nu}(t,{\bf y})]^{-}=[\pi^{\mu}(t,{\bf x}),\pi^{\nu}(t,{\bf y}) ]^{-}
=0 \label{eq:SSTB}
\end{equation} 
\begin{equation}
[\xi_{\mu}(t,{\bf x}),\pi^{\nu}(t,{\bf y})]^{-}=\delta({\bf x}-{\bf y}) \delta_{\mu}^{\nu} \nonumber
\end{equation}
for $2j$=even, and
\begin{equation}
[ \xi_{\mu}(t,{\bf x}),\xi_{\nu}(t,{\bf y}) ]^{+}=[\pi^{\mu}(t,{\bf x}),\pi^{\nu}(t,{\bf y})]^{+}
=0 \label{eq:SSTF}
\end{equation} 
\begin{equation}
[ \xi_{\mu}(t,{\bf x}),\pi^{\nu}(t,{\bf y}) ]^{+}=-\delta({\bf x}-{\bf y}) \delta_{\mu}^{\nu} 
\nonumber
\end{equation}
for $2j$=odd.

\section{Comments} \label{section:Comments}

The result just proven may appear somewhat remote from typical problems encountered in applications
of classical field theory.  For one thing, fields in classical physics are generally 
constrained systems.~\cite{Kupershmidt1992}
For another, if we except the special cases of the electromagnetic and gravitational fields,
problems in classical physics involving canonical realizations of a definite,
but otherwise arbitrary, value of dimensionless angular momentum should be uncommon.  
The connection might better be stated:  Fields described by tensorial canonical realizations
obey fundamental Poisson bracket relations, while fields described by spinorial realizations
obey fundamental Poisson antibracket relations.

While the approach taken in this paper is
Poincar\'{e}-invariant, the treatment is not manifestly covariant, in that it singles out
spacelike slices.  This difficulty is a familiar one in Hamiltonian treatments of problems with
relativistic symmetry.  It appears that the method of proof used in this paper could be cast 
in nonrelativistic language without major change.  The main difficulty would appear
to be replicating the classification of irreducible realizations for general fields.  For the special
case of the classification of bound states of the hydrogen atom, it is possible to recapitulate the 
construction in Section 
\ref{sect3c} in nonrelativistic terms by substituting the Lenz vector for the boost generator 
${\bf K}$,~\cite{Biedenharn1961} but this method is not available in general.  It hardly seems 
necessary, however, to replicate every detail of the structure of irreducible realizations of the 
Poincar\'{e} group in a nonrelativistic treatment, so long as the rotation
group is realized faithfully.  The distinguished role of spacelike slices in the canonical formalism
naturally poses no difficulty in a nonrelativistic setting.

However, it must be questioned whether such a "nonrelatvistic" proof could really be considered 
satisfactory.  Elements of the present demonstration, such as equal-time commutativity of fields, 
the effect of space inversion,
and rotational symmetry, that all follow from the single requirement of Poincar\'{e}
invariance, would evidently enter a nonrelativistic version of the proof as distinct hypotheses. 
This hardly seems parsimonious.  Moreover, it has been argued in a critique of proofs of 
the spin-statistics connection in nonrelativistc quantum mechanics that no nonrelativistic analog 
of local commutativity exists.~\cite{AllenMondragon2003}  Even if this objection be set aside, it 
does not seem that any real advantage is to be gained from a nonrelativistic formulation.

\section{Conclusion}

Simple arguments based upon a field-theoretical canonical treatment of rotational and 
space-inversion symmetry lead
to a proof of the spin-statistics connection for classical Grassman fields which are irreducible 
canonical realizations of the Poincar\'{e} group.


\begin{thebibliography}{5}
\bibitem{footnote1}This condition is presumably not necessary for the existence of a spin-statistics
connection for a classical field, but it greatly simplifies any discussion of the effects of
space inversion.
\end{thebibliography}
\end{document}